\documentclass[aps,prb,twocolumn,groupedaddress,showpacs]{revtex4}

\usepackage{amsmath}
\usepackage{graphicx}

\begin{document}

\title{Heat Bath Approach to Landau Damping and Pomeranchuk Quantum Critical Points}

\author{Johan Nilsson}
\author{A. H. Castro Neto}

\affiliation{Department of Physics, Boston University, Boston, 
Massachusetts 02215}

\date{\today}

\begin{abstract} 
We study the problem of the damping of collective modes close to
a Pomeranchuk quantum critical point in a Fermi liquid. 
In analogy with problems in dissipative open quantum systems, 
we derive the Landau damping of a Fermi liquid by integrating
out a macroscopic number of degrees of freedom from a generating
functional.  Being a reformulation of the linearized Boltzmann equation
this approach reproduces well-known results from the theory of Fermi
liquids. We also study the Bethe-Salpeter equations within the Landau theory and discuss the implications of these results on
quantum phase transitions of the Pomeranchuk type and
its dynamical exponent, $z$. We apply our results to the electronic nematic instability
and find $z=3$ in the collisionless limit.
\end{abstract}


\newcommand{\PD}[2]{\frac{\partial#1}{\partial#2}}
\newcommand{\PDN}[3]{\frac{\partial^{#3}{#1}}{\partial{#2}^{#3}}}

\maketitle

\section{Introduction}
\begin{figure}[htb]
\includegraphics[scale=0.45]{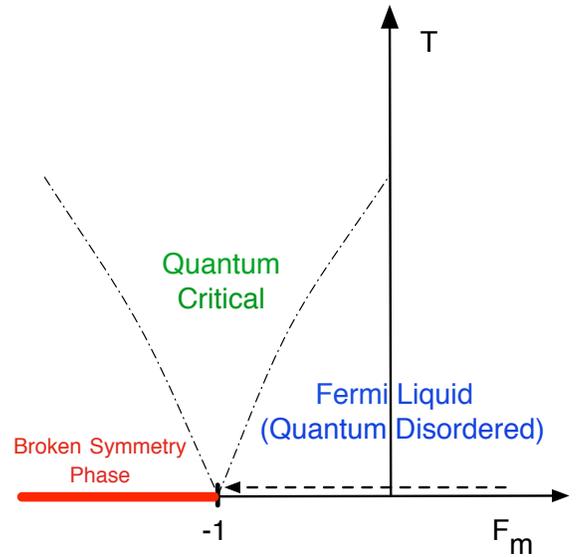}
\caption{Phase diagram of a quantum critical Pomeranchuk instability of order $m$.
Vertical axis: temperature $T$; Horizontal axis: Landau parameter $F_m$.  The dashed arrow line shows
the direction of approach to the quantum critical point used in this work. }
\label{phase_diagram}
\end{figure}

Recently there has been interest in the so-called quantum critical 
Pomeranchuk instability where the interactions in a fermion system are strong
enough to make the Fermi surface soft and ultimately
unstable toward a spontaneous deformation \cite{pomeranchuk}. 
These instabilities have been proposed to be behind 
exotic electronic phases such as electronic liquid crystals in cuprate superconductors \cite{eduardonematic,barci}
and fractional quantum Hall effect \cite{qhe_1,qhe_2}, metamagnetic transitions in electronic systems such as Sr$_3$Ru$_2$O$_7$  \cite{honerkamp}, electronic systems 
close to van Hove singularities \cite{wegner}, 
"hidden order" in URu$_2$Si$_2$ \cite{varma,varma_s}, a mechanism
for generation of spin-orbit coupling \cite{congjun}, and even 
phase transitions in interacting quantum dots \cite{murthy}. 
The Pomeranchuk instability can take place both on a lattice and in the
continuum \cite{eduardonematic,metznersoft,kunyang} and is driven
by attractive quasiparticle-quasiparticle interactions. These interactions
in Fermi liquid theory are parameterized by the Landau parameters, $F^{s,a}_m$,
where $s$ ($a$) stands for the symmetric (anti-symmetric) spin channel and $m = 0,1,...$
for the angular momentum (we  focus on the case of the two-dimensional isotropic
liquid although the three-dimensional case is completely analogous). Pomeranchuk showed
that if some $F^{s,a}_m$ becomes too attractive ($F^{s,a}_m < -1$ in two dimensions)
at $T=0$ then the Fermi surface becomes unstable and cannot sustain collective oscillations.
The simplest example of such a transition is a ferromagnetic (or Stoner) 
instability that occurs when $F^a_0<-1$,  
which is characterized by a diverging magnetic susceptibility, $\chi$ ($\chi \propto 1/(1+F^a_0)$)
A similar instability with a diverging compressibility, ${\cal K}$, can be achieved
by pulling on a liquid to put it into a negative pressure
region when $F^s_0 < -1$ (${\cal K} \propto 1/(1+F^s_0)$) \cite{balibarreview}.
There is also the possibility that the instability takes place in a higher
angular momentum channel. For example, a deformation into 
an elliptical Fermi surface can be achieved by having a strong enough attraction in the $d$-channel,
that is,  $F^s_2 < -1$. The broken symmetry phase in this case has been proposed to be an \textit{electronic nematic}
because of the nomenclature in the classical theory of liquid crystals where an analogous phase exists.
The properties of the broken symmetry phase has been found to be very peculiar with marked non-Fermi liquid properties \cite{eduardonematic}. Although Fermi liquid theory cannot predict what is the nature
of the broken symmetry phase, it can be used to study the approach to the instability from the 
Fermi liquid (quantum disordered) side of the transition (see Fig.\ref{phase_diagram}).

One of the main characteristics of a Pomeranchuk quantum critical point is that
in its proximity  the excitations are not of the simple quasiparticle type but rather
become heavily damped because of the soft nature of the Fermi surface. 
This is so because the Fermi surface allows for a diverging number of low
energy particle-hole pairs to accompany an excited quasiparticle. 
Presumably, a similar situation holds for most of the single-particle excitations 
in the broken symmetry phase \cite{eduardonematic} although the Fermi liquid
theory is not able to access that phase directly. 

A natural way to study the collective modes of an interacting electron
fluid is via Fermi surface bosonization.  The method of bosonization has been very successful in providing a
framework for describing the breakdown of the normal Fermi liquid in one-dimensional (1D) systems, where
instead the Luttinger liquid emerges as the ``standard'' model \cite{voitlong}. In Luttinger
liquids there are no well defined quasiparticles and the electron breaks down into charge and
spin collective modes.  It is now well understood that the anomalous behavior in one dimension is mainly
due to the restricted phase space. Indeed, there exists Pomeranchuk instabilities also in 1D which  
lead to a breakdown of the Luttinger liquid. What happens is that either the charge- or spin- velocity vanishes at the transition.  Since the charge(spin) susceptibility goes as the inverse of the charge(spin)
velocity \cite{voitlong} the signature of the instability is a diverging susceptibility, just as in the higher dimensional case. Thus, the Pomeranchuk instability is a general concept, independent of the dimensionality of
the system under interest.  The higher dimensional bosonization approach was developed many years
ago \cite{haldane,marston_1,marstonreview,antonioboson1,antonioboson2,kopietz}  and reproduces the known results from the theory of Fermi liquids, especially the physics of collective modes and particle-hole excitations which are
described in terms of boson coherent states. The basic reason for this is that both theories describes fermion
systems dominated by forward scattering correctly, and this is what the Landau Fermi liquid is all about \cite{shankarrg}. 

For problems such as those involving classical limits it is a good idea to use a description in terms of
bosons since they can have classical analogs whereas fermions cannot. In particular, when one is studying phase transitions in the language of a Landau-Ginzburg expansion the order parameter is a classical construction
which in the quantum language should be described by bosons.
One can generally trade a fermion system with density-density interactions to one involving bosons by using a Hubbard-Stratanovich transformation and then integrate out the fermions from the path integral that describes the system. This is the standard device to study quantum critical theories in Fermi systems \cite{Hertz,Continentino,Millis}. In such approaches to quantum critical phenomena dissipative terms are generated in the order parameter effective action, the so-called
Landau damping. However, there have been some arguments about whether it is viable to integrate out all of the fermions from the system in this way, or if some properties of the system is lost due to the gapless nature of
the electrons in a Fermi liquid. In particular there are particle-hole-like excitations with arbitrarily low energies which must be treated with care since they may control the critical behavior of a Fermi system. Nevertheless,
the integration of high energy degrees of freedom as one usually does in renormalization group (RG) 
approaches should give sensible results.
Obviously, the particle-hole continuum type excitations do not disappear
in a faithful bosonic reformulation of the system, although their  nature might be disguised.
When some of these modes are 
in resonance with the collective excitations the result is Landau
damping. This occurs only for modes with a velocity smaller than the
Fermi velocity.

In this paper we show that the Landau damping problem can be recast in
terms of a trace over bosonic modes.
In particular the modes associated with the high angular
momentum oscillations of the Fermi surface are traced out since the
collective excitations we are interested are concentrated in the low angular
momentum modes.
Furthermore, using the Bethe-Salpeter equation in the context of the collision integral, we can
study the lifetime of quasiparticles and collective modes in a Fermi liquid close to a Pomeranchuk quantum critical
point. Moreover, we discuss the dynamical critical exponent of the
same quantum critical point.
The paper is organized as follows.
In Section \ref{landaudamp} we show how Landau damping as
described above can be reproduced within a simple purely bosonic reformulation of the
Fermi liquid theory. Here the damping comes about from the coupling to
a bath of harmonic oscillators similarly to what happens in dissipative
quantum mechanics. In Section \ref{collisionintegral} we study the collision
integral in the language of Fermi liquid theory, focusing on the
behavior near the critical point. 
This turns out to be important since collisions sets the lifetime of the
quasiparticles, and as one approaches the critical point the lifetime
decays rapidly. We conclude in Section \ref{conclusions} with a discussion of the
implications of these phenomena on the theory close to the critical
point. In an Appendix we present some mathematical details.

\section{Landau damping from a Boson bath}
\label{landaudamp}

The phenomenon of Landau damping is conventionally thought of as
the damping of a collective mode due
to `a \textit{coherent} interaction of the collective mode with those
particles which ``surf-ride'' on the crests of the running 
wave.'\cite{pinesnoz} 
In this section we show explicitly how Landau damping can also be
viewed as the dissipation of a mode due to its coupling 
to a set of independent oscillators, i.e. a  ``boson bath''. 
Here the origin of these bosons are
particle-hole excitations of the fermion system itself. 
This approach is similar in spirit to theories of dissipation in
quantum mechanics.\cite{feynman,caldeiralegget}

Let us take as our starting point the Landau kinetic equation
that describes the evolution of the quasiparticle distribution
function in the spin symmetric channel, $n_{\vec{p}}(\vec{r},t)$ \cite{pinesnoz}: 
\begin{multline}
  \label{landaukin_n}
  \PD{n_{\vec{p}}(\vec{r},t)}{t} +
\nabla_p \epsilon_{\vec{p}}(\vec{r},t) \cdot \nabla_r
n_{\vec{p}}(\vec{r},t) 
\\
 -
\nabla_r \epsilon_{\vec{p}}(\vec{r},t) \cdot \nabla_p n_{\vec{p}}(\vec{r},t)
= I[n_{\vec{p}'}] \, ,
\end{multline}
where $\epsilon_{\vec{p}}(\vec{r},t)$ is the quasiparticle energy and 
$I[n_{\vec{p}'}]$ is the collision integral ($\vec{p}$ is the quasiparticle momentum,
$\vec{r}$ is the position in real space, and $t$ is time) \cite{baymp}.
 Since the physics goes on close to the Fermi surface it is convenient to
change variables to
$\nu_{\vec{p}}(\vec{r},t)$ which we define by writing the deviation
from the ground state values of quasiparticle distribution as
 $\delta n_{\vec{p}}(\vec{r},t) =
-\bigl( \PD{n_{p}^0}{\epsilon_p} \bigr) \nu_{p}(\vec{r},t)$.
Physically $\nu_{p}(\vec{r},t)$ measure the local deformation of the Fermi surface.
Upon a Fourier transform in space
the linearized kinetic equation for $\nu_{p}(\vec{r},t)$ becomes
\begin{equation}
  \label{landaukin_nu}
    \PD{\nu_{\vec{p}}(\vec{q},t)}{t} + 
i \vec{v}_{\vec{p}} \cdot \vec{q} \;
\bigl[ \nu_{\vec{p}}(\vec{q},t) + \delta \epsilon_{\vec{p}}(\vec{q},t) \bigr]
= I'[\nu_{\vec{p}'}],
\end{equation}
where $\vec{v}_{\vec{p}}$ is the Fermi velocity
at the momentum $\vec{p}$. Again the notation 
\begin{eqnarray}
\delta \epsilon_{\vec{p}} = 
\sum_{\vec{p}'}f_{\vec{p}\vec{p}'}\delta n_{\vec{p}'} \, ,
\end{eqnarray} is standard and
describes the interaction (parametrized by $f_{\vec{p}\vec{p}'}$) among excited quasiparticles.
For an isotropic liquid in two dimensions,  
 $\vec{v}_{\vec{p}}$ is parallel to $\vec{p}$,
and since the magnitude of $p$ is nearly equal to the Fermi momentum
$p_F$ we can
label $\vec{p}$ by an angle
$\theta_{p}$ (measured from the direction of $\vec{q}$).
Then, the interaction $f_{\vec{p}\vec{p}'}$ 
only depends on
the angle between the momenta and is
convenient to expand in Fourier components because of the circular symmetry:
\begin{eqnarray}
  \label{nudef}
  &   \nu (\theta) = \sum_{m} \nu_m e^{i m \theta }  \, ,
  \\
  &   f(\theta,\theta') = \sum_m f_m e^{i m (\theta -\theta') },
\end{eqnarray}
where $f_m = f_{-m}$. 
The sum over $m$ should be cut-off at some integer of order
$E_F / \Lambda \equiv N_{\Lambda}\gg 1$, where $E_F$ is the Fermi energy and
$\Lambda$ is some energy scale below
 which the Landau theory is valid.
Projecting onto the Fourier components and setting the collision
integral to zero for the time being we arrive at
\begin{multline}
  \label{eomlin}
      \PD{\nu_m(\vec{q},t)}{t} +\frac{i v_F q}{2} \Bigl[
       (1+F_{m-1}) \nu_{m-1}(\vec{q},t)  \\ +
       (1+F_{m+1}) \nu_{m+1}(\vec{q},t)
       \Bigr] = 0.
\end{multline}
Here $v_F$ is the Fermi velocity and the $F_n = N(0)f_n$ are the
interaction
parameters
normalized with the density of states at the Fermi energy, $N(0)$.
Note that these equations are linear in the time derivatives and 
couple time derivatives of odd components to the even components
and vice versa. Thus the equations respect time reversal invariance
implying non-dissipative dynamics.

To make connection with the more familiar simple oscillator we eliminate the
odd components in favor of a second time derivative \cite{note}. 
With the notations:
\begin{eqnarray}
    a_m (\vec{q},t) &=& \sqrt{1 +F_m} \, \, \nu_m (\vec{q},t) \, ,
\\
   A_m  &=& \frac{1}{4} (1+F_m) (2 + F_{m-1} + F_{m+1}) \, ,
\\
   B_m  &=& \frac{1}{4} \sqrt{1+F_{m-1}} (1+ F_{m}) \sqrt{1 + F_{m+1}} \, ,
\end{eqnarray}
the equations of motion become:
\begin{multline}
  \label{eomsec}
    \PDN{a_m(\vec{q},t)}{t}{2} +(v_F q)^2 \Bigl[
       A_m a_{m}(\vec{q},t)   \\ +
       B_{m-1} a_{m-2}(\vec{q},t) +
       B_{m+1} a_{m+2}(\vec{q},t)
       \Bigr] =0.
\end{multline}
These equations of motion are also generated from 
Euler-Lagrange equations and the classical Lagrangian density:
\begin{multline}
  \label{lagrangian1}
  \mathcal{L} = \sum_m \PD{a_{2 m}^* (\vec{q},t)}{t}
  \PD{a_{2 m} (\vec{q},t)}{t}  \\
  - (v_F q)^2 \sum_{m,n} a_{2 m}^* (\vec{q},t) V_{m n} a_{2 n} (\vec{q},t),
\end{multline}
where 
\begin{eqnarray}
V_{m n} = \delta_{m,n} A_{2 m} + \delta_{m-1,n}B_{2 m -1} +
\delta_{m+1,n} B_{2 m +1} \, .
\label{vmn}
\end{eqnarray}
Notice that (\ref{lagrangian1}) describes a set of coupled harmonic
oscillators coupled by "springs" with spring constants given by $V_{m,n}$.
This result shows the
clear collective nature of the Fermi surface modes.

We can also encode the evolution of the system with
the methods of path integrals\cite{negeleo} in the real time partition
function:
\begin{equation}
  \label{Zdef}
  Z = \int Da^* Da \, \, e^{i S},
\end{equation}
where we also perform a Fourier transform in time
\begin{equation}
  \label{action1}
  S = \sum_{\vec{q},m,n} \int d\omega \, a_{2m}^{*}(\vec{q},\omega) \, 
   \tilde{V}_{mn} \, a_{2n}(\vec{q},\omega),
\end{equation}
with,
\begin{eqnarray}
\tilde{V}_{mn} &=& (v_F q)^2 \left[s^2 \delta_{mn} -V_{mn}\right] \, ,
\label{tvmn}
\\
s&=& \omega/(v_F q) \, .
\end{eqnarray}
It is clear from these results that the propagator of the collective modes is 
essentially given by the inverse of the coupling matrix $\tilde{V}_{m,n}$, that is,
$\tilde{V}_{m,n}^{-1}$. Therefore, the eigenmodes of $\tilde{V}$ are the
collective excitations of the system and its eigenenergies determine the frequency
of oscillations of these modes.

We can gain further insight into the problem by splitting the action (\ref{action1}) 
into two decoupled sectors ($S =S^+ +S^-$) by using the symmetric and antisymmetric
combinations  
\begin{eqnarray}
a_{m}^{\pm} = \frac{1}{\sqrt{2}} (a_{m} \pm a_{-m} ) \, , 
\end{eqnarray}
defined for $m \geq 0$.
These corresponds to the longitudinal ($+$) and transverse ($-$) modes in the sense
of having even/odd parity with respect to the line 
 defined by $\vec{q}$.
Notice that the transverse modes have no density (i.e. $m=0$) component.

Formally, there can be an infinite number of Landau parameters $F_m$
that parameterize the interactions in a Fermi liquid. In practice, however,
only a few Landau parameters are taken into account \cite{baymp}. In the
case of the Pomeranchuk instability we can concentrate on a single 
Landau parameter, that is, the most singular, since all the other parameters
 only provide trivial renormalizations of the Fermi liquid properties. 
Without lack of generality let us assume that there are $M$ Landau parameters
in the description of a quantum liquid, that is,  the $F_m$ are zero for 
$m \geq M+1$. In this case the matrix $\tilde{V}_{m,n}$ in (\ref{tvmn}) 
for $m,n \geq M+1$ (we call it the high-$M$ block) can be
written in the longitudinal and transverse sectors in terms of blocks of the form:
\begin{equation}
  \label{blockmatrix}
  \begin{pmatrix}
   s^2 - \frac{1}{2} & -\frac{1}{4} & 0 & 0 & \cdots \\
   -\frac{1}{4} & s^2 -\frac{1}{2} & -\frac{1}{4} & 0 & \\
   0 & -\frac{1}{4} & s^2 -\frac{1}{2} & -\frac{1}{4} &  \\
   0 & 0 & -\frac{1}{4} & s^2 -\frac{1}{2} &  \\
   \vdots &  &  &  & \ddots
  \end{pmatrix}.  
\end{equation}
The matrix (\ref{blockmatrix}) (assumed to have size $N \times N$)
have orthonormal eigenvectors of the form:
\begin{equation}
  \label{eigenvectors1}
    \vec{\varphi}(k) = \sqrt{\frac{2}{N+1}} \begin{pmatrix} 
\sin(k), \sin(2k) , \sin(3k) ,\ldots , \sin(N k) 
\end{pmatrix}^T,
\end{equation}
with respective 
eigenvalues $\lambda(s,k) = s^2 - (1+\cos(k))/2$ (the allowed values for the
$k$'s are: $k = \pi n / (N+1)$ where $n=1 \ldots N$).
The block (\ref{blockmatrix}) is coupled to the mode $M$ via two off-diagonal terms
in $V_{m,n}$. Schematically we have:
\begin{equation}
  \begin{pmatrix}
    \mbox{low-M block} & 
    \begin{matrix}
      \vdots &  \vdots &  \\ 
      0 & 0 & \cdots \\
      B_{2M+1} & 0 & \cdots
    \end{matrix}
\\
    \begin{matrix}
       \cdots & 0 & B_{2M+1} \\ 
       \cdots & 0 & 0 \\
       & \vdots & \vdots
    \end{matrix}    
& \mbox{high-M block}
  \end{pmatrix}.
\end{equation}
Using the diagonalized form of the block matrix it is not
hard to integrate out the high-$M$ modes. The end result is to change
\begin{equation}
  \label{eq:dA}
     A_M \rightarrow A_M + \delta A_M,
\end{equation}
in the low-$M$ block of $\tilde{V}_{mn}$, where
 \begin{eqnarray}
 \delta A_M = 2 B_{2 M +1}^2 G(s)
 \end{eqnarray}
 and 
\begin{equation}
\label{Gdef}
G(s) = \frac{1}{N+1} \sum_k \frac{\sin^2(k)}{s^2 - (1+\cos(k))/2}.
\end{equation}
From this result we see that $G(s)$ as a function of complex $s$ is analytic
in the upper and lower half plane. It has a set of poles on the real
axis between $s=-1$ and $s=1$. In the limit of $N \rightarrow \infty$
the poles become dense resulting in a branch cut. By taking $N\rightarrow \infty$,
$G(s)$ can be expressed as an integral:
\begin{equation}
  \label{Geval}
  G(s) = \int_{-\pi}^{\pi}\frac{dk}{\pi} \frac{\sin^2(k)}{2 s^2 - 1 - \cos(k)}.
\end{equation}
Taking $s$ to lie away from the branch cut one can turn the integral
into a contour integral around the unit circle centered at the
origin. The resulting integral is easily evaluated for $|s|>1$ by
the residue theorem and can be
analytically continued to $|s|<1$ with the result:
\begin{equation}
  \label{Gval}
  G(s)= -2 + 4 s^2 - 4 \sqrt{s^4 -s^2}.
\end{equation}
So that for real $s$ and $|s|>1$
\begin{equation}
  \label{Glarge}
  G(s) = -2 + 4s^2 -4 |s|\sqrt{s^2-1},
\end{equation}
while for real $s$ and $|s|<1$
\begin{equation}
  \label{Gsmall}
  G(s \pm i0^+) = -2+ 4s^2 \mp 4 i s\sqrt{1 - s^2}. 
\end{equation}
Hence, the final result of integrating out the high angular momentum modes
is the introduction of a shift in the Landau parameters with the introduction
of an imaginary part when $|s|=|\omega|/(q v_F)<1$, which is nothing
but the Landau damping. By only keeping the low-$M$ modes the 
time reversal invariance has been broken.

Since what we have done is a reformulation of the Landau theory it
should give the same result as the conventional approach. A simple
check amounts to
taking only $F_0 \neq 0$ and integrate out all modes except the
zeroth one in the longitudinal sector, that is, $M=0$. 
The well-known zero sound mode should then come out as the zero
of the only remaining matrix element (which is equivalent to having a
pole in the propagator). Explicitly we obtain:
\begin{equation}
  \label{ZScheck}
  \frac{s^2}{2}-\frac{A_0+2 \delta A_0}{2} = 0.
\end{equation}
Substituting the expression for $A_0$, $B_1$ and $G(s)$ we get an
undamped solution for $F_0 > 0$ with
\begin{equation}
  \label{ZScheck2}
  |s| = \frac{1+F_0}{\sqrt{(1+F_0)^2 - F_{0}^2}},
\end{equation}
which is the dispersion of the zero mode in two dimensions, see e.g. Ref.[\onlinecite{antonioboson1}].

\section{Lifetime effects close to a Pomeranchuck quantum critical point}
\label{collisionintegral}

In this section we investigate the effect of the proximity of a Pomeranchuk
quantum critical point on the single particle and collective modes of a
Fermi liquid.  Using the collision integral formalism we study the quasiparticle
lifetime via the Bethe-Salpeter equation close to the quantum critical point.
We show that the critical point associated with the quantum fluctuations
change substantially the quasiparticle lifetime without affecting the damping
of the collective modes at low angular momentum.

\subsection{Single particle states}

We follow the usual procedure and study the
lowest order process, i.e. quasiparticle-quasiparticle scattering.
Then the expression for the collision integral becomes \cite{baymp}:
\begin{eqnarray}
  \label{CIB}
  I[n_{\vec{p}}] = 2\pi \sum_{\vec{p}_2 \sigma_2} 
\sum_{\stackrel{\vec{p}_3 \sigma_3}{\vec{p}_4 \sigma_4}}'
  |<3 4|t|1 2>|^2 \nonumber \\ 
\delta_{\vec{p}_1+\vec{p}_2 ,\vec{p}_3+\vec{p}_4} 
  \delta_{\sigma_1+\sigma_2, \sigma_3 +\sigma_4}
\delta(\epsilon_1 + \epsilon_2 -\epsilon_3 -\epsilon_4) \nonumber \\
\Bigl[ n_3 n_4(1-n_2)(1-n_1) - n_1 n_2 (1-n_3) (1-n_4) \Bigr],
\end{eqnarray}
which describe scattering 
processes in and out of the state $\vec{p}$ which is
labeled $1$. The spin information can be encoded into the
the definition of  the matrix element $|<t>|^2$ 
as described in Ref. [\onlinecite{baymp}], so we
drop the spin information from now on (or assume a
spin-independent scattering amplitude).

To make progress we linearize the collision integral
 in the deviations $\delta n_i$ from (local) equilibrium $n_{i}^{0}$. 
Moreover, we  employ the relaxation
time approximation which amounts to keeping only the term proportional to
$\delta n_1$:
\begin{equation}
  \label{eq:4}
  I = -\frac{1}{\tau_{\vec{p}}} \delta n_{\vec{p}},
\end{equation}
where
\begin{eqnarray}
  \label{eq:5}
  \frac{1}{\tau_{\vec{p}}} = 2\pi \sum_{\vec{p}_2} 
\sum_{\vec{p}_3,\vec{p}_4}'
  |<3 4|t|1 2>|^2 \nonumber \\ 
\delta_{\vec{p}_1+\vec{p}_2 ,\vec{p}_3+\vec{p}_4} 
\delta(\epsilon_1 + \epsilon_2 -\epsilon_3 -\epsilon_4) \nonumber \\
\Bigl[ n_{3}^0 n_{4}^0(1-n_{2}^0) + n_{2}^0 (1-n_{3}^0) (1-n_{4}^0) \Bigr].
\end{eqnarray}
One can do better than the relaxation time approximation
but it is not necessary for our purposes \cite{baymp} .

Now we introduce the momentum transfer $\vec{q}$, and the
energy transfer $\omega$ via the identity 
$\delta(\epsilon_1 + \epsilon_2 -\epsilon_3 -\epsilon_4) =
\int d\omega \delta(\epsilon_1 - \epsilon_3 - \omega)
\delta(\epsilon_2 - \epsilon_4 + \omega)$.
We  also specialize 
to nearly forward scattering. This is an approximation in
three dimensions where it amounts to neglecting out of plane
scattering. In two dimensions it is essentially exact because of momentum
conservation and the reduced phase space. This means that we  restrict
the sum over $\vec{q}$ to $q < q_c$, where $q_c \ll p_F$ is some
cut-off momentum. Dropping the superscripts on the occupation numbers for clarity we then get
\begin{eqnarray}
  \label{lt3}
  \frac{1}{\tau} = 
     \frac{2 \pi}{\left[1-n(\epsilon)\right]} \int d\omega
     \sum_{\vec{q}} \sum_{\vec{p}'}
  |<3 4|t|1 2>|^2 \nonumber \\ 
\delta(v_F q \cos(\theta_p) - \omega) 
\delta (v_F q \cos(\theta_{p'}) - \omega) \nonumber \\
\left[1-n(\epsilon -\omega \right)] n(\epsilon') 
\left[1-n(\epsilon'+\omega)\right].  
\end{eqnarray}
This form is convenient when $\epsilon >0$, for
$\epsilon <0$ one can instead exchange $n \leftrightarrow 1-n$ everywhere.
The end result is even in $\epsilon$.

To proceed we need an expression for the matrix element.
This one can get from the solution of the Bethe-Salpeter equation,
which describes repeated scattering of a quasiparticle-quasihole pair:
\begin{multline}
  \label{bethesalpeter}
  t_{\vec{p} \vec{p}'}(\vec{q},\omega +i\eta) = f_{\vec{p} \vec{p}'} 
\nonumber \\ -
\sum_{\vec{p}'' \neq \vec{p}'} f_{\vec{p} \vec{p}''}
\frac{\vec{q} \cdot \nabla_{p''}n_{p''}^0}{\omega + i\eta - \vec{q}
  \cdot \vec{v}_{\vec{p}''}}
t_{\vec{p}'' \vec{p}'}(\vec{q},\omega +i\eta).
\end{multline}
The formalism for solving the Bethe-Salpeter equation in three dimensions can be
found in e.g. Ref. [\onlinecite{pethickfinite}]. Here we  outline
the procedure in two dimensions.

Firstly one expands $t$ in Fourier components
\begin{equation}
  t_{\vec{p} \vec{p}'}(\vec{q},\omega +i\eta) = 
  \sum_{m, m'} t_{m,m'} (\vec{q},\omega +i\eta) e^{i (m \theta_{p} - m'
  \theta_{p'})},
\end{equation}
where the angles $\theta$ are measured with respect to $\vec{q}$.
A matrix equation for $t_{m,m'}$ can be derived from the orthogonality
between the Fourier components
\begin{equation}
  \label{teq}
  (1+F_m) t_{m,m'} = \delta_{m,m'} f_m +F_m \sum_{n} K(n-m) t_{n,m'},
\end{equation}
where 
\begin{equation}
  K(n) = \int_{0}^{2\pi}\frac{d\theta}{2 \pi} \frac{s}{s-\cos(\theta)}e^{i n \theta},
\end{equation}
and
\begin{equation}
  s=\frac{\omega + i\eta}{v_F q}.
\end{equation}
Explicit expressions for $K(n)$ can be found in
Ref. [\onlinecite{antonioboson1}]. 
Given the values of $\{f_n\}$ one can solve (\ref{teq}) to get $t_{m,m'}$.

Close to a Pomeranchuk quantum critical point when one of the
Landau parameters reaches a critical value we can keep only
the relevant Landau parameter (see Fig.\ref{phase_diagram}).
In this paper we  study
the ``nematic'' critical point and choose to keep only $F_2=F_{-2}$
nonzero in which case it is straightforward to solve for $t$. The only
nonzero components are
\begin{eqnarray}
t_{2,2}  &=& t_{-2,-2} = \frac{ 1+F_2 - F_2 K(0) }{D} f_2 \, ,
 \nonumber
 \\
t_{2,-2} &=& t_{-2,2} = \frac{F_2 K(4)}{D} f_2 \, ,
\end{eqnarray}
where 
\begin{eqnarray}
D=[1+F_2 -F_2 K(0)]^2 - [F_2 K(4)]^2 \, .
\end{eqnarray}

Now that we have an explicit expression for $<t>$ we can go back to the
expression for the lifetime in (\ref{lt3}), turn the sums into
integrals and perform the
angular integrals with the result
\begin{eqnarray}
  \label{lt4}
  \frac{1}{\tau} = 
     \frac{N(0) L^2}{(2\pi v_F)^2 \left[1-n(\epsilon)\right]} \int d\omega 
\left[1-n(\epsilon -\omega \right)] \nonumber
\\ \times \int d\epsilon' n(\epsilon') 
\left[1-n(\epsilon'+\omega)\right] 
 \int_{|\omega|/ v_F}^{q_c} \frac{dq}{q}  |t(s,q)|^2,
\end{eqnarray}
where
\begin{eqnarray}
  \label{t2}
  |t(s,q)|^2 = \frac{16}{1-s^2}\bigl[1-8 s^2(1-s^2)\bigr]|t_{2,2}+t_{2,-2}|^2
\nonumber \\
 +512 s^4 (1-s^2) \bigl( |t_{2,2}|^2 + |t_{2,-2}|^2 \bigr).
\end{eqnarray}
The first term in (\ref{t2})
gives the dominating contribution (from small $s$)
so we  only keep it in what follows. Notice that due to the
integration limits on $q$ in (\ref{lt4}) we have 
$s \leq 1$ and the singularity near $s=1$ is
removed by the behavior of $|t_{2,2}+t_{2,-2}|^2$.

Usually, i.e. for not too strong attractive interaction, 
it is a good approximation to take $t$ to only depend on $s$. 
Close to the critical point however, one must also
include a $q$-dependence in $t$, explicitly 
we take 
\begin{eqnarray}
1+F_2 = \delta +\kappa (q / q_c)^2 \, ,
\end{eqnarray}
where $\kappa \propto \partial^2 F_2/\partial q^2$.
Now we can trade the integral over $q$ to one over $s$.
The integral is then dominated by the small-$s$ contribution and it is
possible to approximate the integral by a scaling argument. 
For small $s$ we have:
\begin{eqnarray}
  \label{tapprox}
 t_{2,2}+t_{2,-2} &=& \frac{f_2}{1+F_2 -F_2 [K(0)+K(4)]}  \, ,
 \nonumber 
 \\ 
 &\approx&  \frac{f_2}{\delta +\kappa (q/q_c)^2 +2 i F_2 s} \, ,
\nonumber
\\
&\approx& 
\frac{f_ 2 s^2}{\delta s^2 + (\omega / \omega_c)^2 -2i s^3}.
\end{eqnarray}
Using (\ref{tapprox}) the $q$-integral in (\ref{lt4}) can be written as:
\begin{equation}
  \label{eq:8}
  16 |f_2|^2 \int_{|\omega|/v_F q_c}^1 ds 
     \frac{s^3}{[\delta s^2 + (\omega / \omega_c)^2]^2 + 4 s^6}.
\end{equation}
For $\delta \to 0$ (that is, at the quantum critical point) 
we scale $s^3 \propto \omega^2$ so that the integral becomes
$\propto \omega^{-4/3}$.
Performing the remaining integrals at $T=0$ one get
\begin{equation}
  \label{eq:tau1}
  \frac{1}{\tau} \propto |\epsilon|^{2/3} \, ,
\end{equation}
leading to a very anomalous energy dependence of the quasiparticle
lifetime \cite{eduardonematic,metznersoft}. For finite $\delta$ (away from the quantum critical point, inside of the quantum
disordered regime) we should recover the standard Fermi liquid
result. By scaling $s \propto \omega$, we get: 
\begin{equation}
  \label{eq:tau2}
  \frac{1}{\tau} \propto \epsilon^2 \ln (\omega_c / \epsilon)^2 + 
\mbox{const} \times \epsilon^2,
\end{equation}
which is the Fermi liquid result \cite{metznersoft}. 
The above results can be summarized by the scaling form of the quasiparticle lifetime
that can be obtained using  the contour integral trick of Pethick \cite{pethickfinite} 
that we give in the appendix. The general result for the quasiparticle
lifetime is: 
\begin{equation}
  \label{eq:tau3}
  \frac{1}{\tau} = \mbox{const} \times \epsilon^2 + 2 \omega_c^2 (\epsilon/\omega_c)^{2/3}
F\Bigl[\Bigl( \frac{3}{\delta}\Bigl)^3 \Bigl(\frac{\epsilon}{\omega_c}\Bigr)^2\Bigr],
\end{equation}
where $F(\xi)$ is a scaling function such that $F(\xi) \approx 1$ for $\xi \gg 1$
and $F(\xi) \approx -\xi^{2/3}\log(\xi)$ for $\xi \ll 1$, reproducing the results of eqs.
(\ref{eq:tau1}) and (\ref{eq:tau2}).   
A similar crossover formula (for the imaginary part of the
self-energy) in the case of a ferromagnetic 
quantum critical point was found by Chubukov \textit{et al}.\cite{chubukovself,chubukov3}.  
Notice that our results can be generalized for any kind of Pomeranchuk quantum critical
point given the critical Landau parameter. 
The result of this lowest order approximation is thus that close to the ``nematic'' 
critical point in 2D the lifetime of the quasiparticle dominates over
the energy and the concept of the usual quasiparticle is not well-defined.
To see the real fate of the quasiparticle one must include higher
order scattering processes or use a more microscopic theory. 
Nevertheless, the calculation in this section signals the breakdown of
normal Fermi liquid behavior close to a Pomeranchuk quantum critical point.

Studies by other groups of similar models also find a breakdown of
Fermi liquid theory at the critical point. The $\epsilon^{2/3}$
behavior in two dimensions
has also been found before in related systems, see
e.g. Ref.
[\onlinecite{altshulerioffemillis2,eduardonematic,metznersoft,chubukovself,chubukov2}].

\subsection{Collective modes}

To calculate the relaxation time for collective modes one must keep
in mind that inside the linearized
collision integral there are terms of the form
$\delta n_{\vec{p}} - \delta n_{\vec{p}+\vec{q}}$ where the second term is
neglected in the usual relaxation time approximation. The existence of
the second term makes collisions less effective in the damping of
collective modes. More explicitly, linearizing the collision integral (\ref{CIB}) 
for the $\nu$'s in (\ref{landaukin_nu}) we get:
\begin{eqnarray}
  \label{Iprime}
  I'[\nu] = -\frac{2\pi}{1-n(\epsilon_1)} \sum_{\vec{p}_2} 
\sum_{\vec{p}_3,\vec{p}_4}'
  |<3 4|t|1 2>|^2 \nonumber \\ 
\delta_{\vec{p}_1+\vec{p}_2 ,\vec{p}_3+\vec{p}_4} 
\delta(\epsilon_1 + \epsilon_2 -\epsilon_3 -\epsilon_4) \nonumber \\
n_{2}(1-n_{3})(1-n_4)\Bigl[\nu_1 - \nu_3 +\nu_2 -\nu_4 \Bigr].
\end{eqnarray}
Expanding the $\nu$'s in Fourier components as in Section
\ref{landaudamp} and projecting the collision integral we get 
\begin{equation}
  \label{eq:14}
  \frac{1}{\tau_m} = m^2 \frac{1}{\tau'}.
\end{equation}
where $1/\tau'$ is given by (\ref{lt4}) with the substitution $\int
dq \rightarrow \int dq \; q^2 /2 p_{F}^2$.
To get this result we have neglected the term $\nu_2 - \nu_4$ since this term is
typically of the same size as the first (i.e. $\nu_1 - \nu_3$)
 and just gives a different
prefactor in the calculation of $1/\tau$ of order $1$.
More important is the expansion of $\cos[m (\theta + q/ p_F)]$ in
powers of $m q /p_F$, which is only valid for
 $m \ll p_F /q_c \approx N_{\Lambda}$.
The terms with odd powers of $q$ vanish upon performing the
angle averages.

For these modes the simple scaling evaluation of the integrals gives:
\begin{equation}
  \label{eq:15}
  \frac{1}{\tau_m} \propto \Bigl(\frac{m}{N_{\Lambda}}\Bigr)^2 
    \epsilon^{4/3},
\end{equation}
for $\delta \to 0$ and
\begin{equation}
  \label{eq:16}
  \frac{1}{\tau_m} \propto \Bigl(\frac{m}{N_{\Lambda}}\Bigr)^2 
    \epsilon^{2},
\end{equation}
for finite $\delta$. From this we can conclude that the collisionless
approximation is good for self-driven modes ($\omega \approx \epsilon$)
when $(m/N_{\Lambda})^2 \ll 1$.
Note that this implies that it makes sense to talk about (e.g.) the
$M=2$ mode also in the critical regime.
The high-$M$ modes are more sensitive to low-$q$ scattering. 
For these it is not possible to relate $\cos[m(\theta +q / p_F)]$ in any simple way by
$\cos(m\theta)$ and hence the
single particle result above should be a good approximation. This
means that the high-$M$ modes are heavily damped in the critical regime. 
For intermediate $M$ we expect a crossover between the two limiting cases.
Also, note that in all cases (single-particle or collective)
there is a prefactor of $1/(k_F v_F) \approx 1/\epsilon_F$ and the correct powers of $\omega_c$ 
to make the dimensions correct. 

\section{Quantum Critical Regime - Discussion}
\label{conclusions}

Recently Yang\cite{kunyang} has
argued in favor of a dynamical critical exponent of $z=2$ for the
Pomeranchuk quantum critical point in 2D. This is in disagreement with
e.g. the results of Ref. [\onlinecite{eduardonematic}].
The claim that the other non-critical modes just 
renormalize the couplings in the critical theory is not obvious.
In particular, the low-energy particle-hole fermion
excitations which are reproduced within the bosonic formulation as
linear combinations of the high angular
components are neglected in Ref.[\onlinecite{kunyang}].
The calculation in Section \ref{landaudamp} illustrates the effect of
these modes in the collisionless regime, and in
a simpler setting than that of multidimensional bosonization which
requires a complicated Bogoliubov transformation to achieve similar
results.\cite{antonioboson2}

Taking only $F_0$, $F_1$ and $F_2 \neq 0$  we can get the equations for the transverse
and longitudinal modes from the result of Section \ref{landaudamp}.
If we integrate out all the modes that  we are not interested in (i.e. $M \geq 3$),
the resulting equations for the modes are:
\begin{equation}
  \label{mode-}
  s_-^2-A_2-\delta A_2 =0,
\end{equation}
for the transverse ($-$) mode and
\begin{equation}
  \label{mode+}
  \begin{vmatrix}
    (s_+^2-A_0)/{2} & -B_1 \\
    -B_1 & s^2_+-A_2-\delta A_2
  \end{vmatrix}
   = 0,
\end{equation}
for the longitudinal ($+$) mode. From these equations one can extract 
the leading scaling behavior 
to be $s_{-}^2 \sim [\delta + \kappa (q/q_c)^2]/4$ and
$s_{+} \sim -i [\delta+\kappa (q/q_c)^2 /2]$, 
assuming that only the $d$-channel
goes critical. So the transverse mode has a dynamical critical exponent 
of $z=2$ and the longitudinal
$z=3$ in agreement with Ref. [\onlinecite{eduardonematic,notenew}]. The
longitudinal mode, being slower, dominates the critical behavior.

As we found in Section \ref{collisionintegral}, the collisionless
behavior is only a good approximation for low-$M$ modes in the
critical regime. The question that arises is what would be the
effect of collisions on the modes that participate in the Landau
damping.
The coupling of the $M=2$ mode to the rest of the system
still results in damped motion.
Generally one can integrate out the modes with $M \geq 3$ with the
same result as above in (\ref{mode-}) and (\ref{mode+}).
The difference is that we no longer have an explicit
expression for $\delta A_2$ since the approximations of the
previous section break down. 
However, we still must have $\delta A_2 = 2B_{3}^2 U(q,\omega)$, 
where $U(q,\omega)$ is a function such that $U(q,0)= U(s=0) = -2$. 
The calculation of $U(q,\omega)$ is beyond of the scope of this
work. Nevertheless, if $U(q,\omega)$ is known, the longitudinal
mode equation becomes:
\begin{equation}
  \label{eq:2}
  s_+^2 -\frac{1}{8} \Bigl[\delta + \kappa
 \Bigl(\frac{q}{q_c}\Bigr)^2\Bigr]
 \times 
\Bigl[ U(q,\omega) - U(q,0) \Bigr] = 0.
\end{equation}
If we assume that $U(q,\omega)-U(q,0) \sim -i \omega^{\alpha} / q$ 
in the $\omega \to 0$ limit, the
dynamical critical exponent of the mode is $z=3/(2-\alpha)$.
This result allows us to put bonds on the value of the dynamical
exponent since we expect $2/3 \leq \alpha \leq 1$ in
such a way that $9/4 \leq z \leq 3$.
One can speculate that the modes that are responsible for the damping
of the critical mode, having very low energy, are weakly damped since there
is not much for them to decay into because of the small phase space. This would
hint towards $z=3$ even when collisions are taken into account.
Another argument in favor of $z=3$ is that 
the collisionless limit is a good approximation when the mean
free path is much longer than the wavelength,
 which translates into $1/\tau
\ll q v_F $.\cite{pinesnoz}
 Taking $1 / \tau \sim \omega^{2/3}$ and $\omega \sim q^3$
the inequality is clearly  satisfied as $q\rightarrow 0$
indicating the validity of the collisionless limit.

A questionable thing with the whole approach close to the critical
point is whether the calculations are self-consistent.
The simplest particle-particle scattering process gave a lifetime
that is proportional to $\epsilon^{2/3}$.
The same result can be obtained from a 1-loop calculation of the
imaginary part of the fermion self energy, which describes similar
physics.
Taking this seriously would invalidate the quasiparticle picture, and
hence make the whole ground for the calculation questionable. Notice,
however, that although the quasiparticle description seems to break down,
the low angular momentum 
collective spectrum remains essentially unchanged (except by the
damping of the collective modes given by the subleading contributions (\ref{eq:15}) and (\ref{eq:16})).

One further circumstantial evidence for a breakdown of the quasiparticle near the
quantum critical point comes from  the theory of multi-dimensional
bosonization\cite{antonioboson2}.
The calculations are tractable for an isotropic interaction, the
connection to the notations in this paper is $Ng = F_0$.
The Bogoliubov transformation
used to diagonalize the system connects different patches $i$ and $l$ of the Fermi
surface according to the matrices:
\begin{eqnarray}
  \label{eq:1}
  \mathcal{M}_{il} = \delta_{il} + \frac{1}{N} \frac{Ng}{1+Ng} 
  \frac{\sqrt{s_i s_l}}{s_l - s_i} \, ,
  \nonumber \\
  \mathcal{N}_{il} = - \frac{1}{N} \frac{Ng}{1+Ng}
  \frac{\sqrt{s_i s_l}}{s_l + s_i} \, ,
\end{eqnarray}
where $s_i = \cos(\theta_i)$. 
Away from the critical point the second term in $\mathcal{M}$ represent the
dressing of the particle by interactions, and $\mathcal{N}$ is always
small. Close to the critical point however, the factor
$Ng/(1+Ng) = F_0/(1+F_0)$ blows up when $F_0 \to -1$ and a broad dressing cloud dominates the
excitations, also $\mathcal{N}$ is no longer negligible and is an important part of
the excitation. Moreover the expression for the quasiparticle residue
$Z= \exp\left\{-\frac{1}{4}(\frac{F_0}{1+F_0})^2\right\}$ 
shows that it vanishes exponentially as the
transition is approached.
The calculation for the $d$-channel Pomeranchuk transition is much more
complicated because the kernel of the integral equation
is no longer a product kernel. We expect similar conclusions however.
Because the Fermi liquid theory breaks down close to the critical
point a calculation within the Fermi liquid theory itself doesn't make much sense there, 
except for indicating its own breakdown.
The correct description is not in terms of 
the standard Fermi liquid theory and
it would be very interesting to know what it is.

In summary, we have presented an alternative approach for the study of Landau damping
in Fermi liquids based on the trace over high angular momentum modes. We find that this
approach reproduces well-known results in the literature and allows for the study of 
the quasiparticles and collective modes close to a Pomeranchuk quantum critical point.
Although our approach is general and can be used to study any instability of the Pomeranchuk
type, we focused on the problem of the electronic nematic that has been discussed in the
literature recently \cite{eduardonematic,kunyang,kee,khavkine,eduardonew}. We found that the quasiparticle lifetime becomes
anomalously short at the quantum critical point leading to a
breakdown of the Fermi liquid description.

\acknowledgments
This work was supported through NSF grant DMR-0343790.

\appendix*
\section{Contour integral}

The formula for the inverse lifetime can be written
(at $T=0$ and $\epsilon>0$ for simplicity)
\begin{multline}
  \label{eq:13}
  \frac{1}{\tau} = 
     \frac{1}{2\pi v_F k_F} \int_0^{\epsilon} d\omega \omega
 \Bigl[-c_{s}^{<} + N^2(0) \int_{0}^{1} \frac{ds}{s}  |t(s,q)|^2 \Bigr]
\\
\equiv 
     \frac{1}{2\pi v_F k_F} \int_0^{\epsilon} d\omega \omega
 (-c_{s}^{<} + I_s).
\end{multline}
To get this we have added and subtracted the contribution of the
$N^2(0)$ times the $s$-integral down to zero, this is approximately
\begin{eqnarray}
  \label{eq:10}
    c_{s}^{<} = 16 \int_0^{|\omega|/v_F q_c} ds 
     \frac{s^3}{[\delta s^2 + (\omega / \omega_c)^2]^2 + 4 s^6}
\nonumber \\
= 16 \int_0^1 dx 
     \frac{x^3}{[\delta x^2 + \kappa]^2 + 4 x^6
     (\frac{\omega}{v_F q_c})^2},
\end{eqnarray}
which to leading order is a constant (for finite $\kappa$), 
note that we set $F_2 =-1$
everywhere where it is not dangerous to do so.

Now we can follow Pethick\cite{pethickfinite} and turn the integral in $I_s$ 
into two contour integrals, the difference being that in our case the
two contours have different orientation
\begin{equation}
  \label{eq:6}
 I_s = -  \frac{2 N(0)}{i}\oint_{c}
  \frac{ds}{s^2}\frac{1}{\sqrt{1-s^2}}
  \frac{1}{1+4s^2(s^2 -1)} (t_{2,2}+t_{2,-2}),
\end{equation}
using the fact that the integral is dominated by small values of $s$ one can
rewrite it as
\begin{equation}
  \label{eq:7}
  I_s = -8 \mathcal{P}\int_{0}^{\infty} dz 
\frac{1}{\delta z^2 - (\omega/\omega_c)^2+2z^3},
\end{equation}
after a deformation of the contour to the imaginary axis. 
Scaling $z = \delta x /6$ we get
\begin{equation}
  \label{eq:11}
  I_s = -16 \bigl(\frac{3}{\delta}\bigr)^2
  \mathcal{P} \int_0^{\infty} dx \frac{1}{x^3 + 3 x^2 - 4 Y},
\end{equation}
where $Y = (3 /\delta )^3 (\omega / \omega_c)^2$.
Finally performing the remaining integral we obtain
\begin{multline}
  \label{eq:12}
  \frac{1}{\tau} = \frac{1}{2\pi v_F k_F}
  \Bigl[ - c_{s}^< \frac{\epsilon^2}{2}
\\+ 2 \epsilon^2 (\frac{\omega_c}{\epsilon})^{4/3} 
  \frac{1}{\xi^{1/3}} 
  \mathcal{P}\int_{1}^{\infty} dx 
        \log \left[\frac{x^3 -3 x + 2(1-2 \xi)}{x^3 - 3x +2} \right] \, , 
\end{multline}
which is of the desired scaling form with 
$\xi = (3 /\delta )^3 (\epsilon / \omega_c)^2$.

\bibliography{damping.bib}

\begin{thebibliography}{40}
\expandafter\ifx\csname natexlab\endcsname\relax\def\natexlab#1{#1}\fi
\expandafter\ifx\csname bibnamefont\endcsname\relax
  \def\bibnamefont#1{#1}\fi
\expandafter\ifx\csname bibfnamefont\endcsname\relax
  \def\bibfnamefont#1{#1}\fi
\expandafter\ifx\csname citenamefont\endcsname\relax
  \def\citenamefont#1{#1}\fi
\expandafter\ifx\csname url\endcsname\relax
  \def\url#1{\texttt{#1}}\fi
\expandafter\ifx\csname urlprefix\endcsname\relax\def\urlprefix{URL }\fi
\providecommand{\bibinfo}[2]{#2}
\providecommand{\eprint}[2][]{\url{#2}}

\bibitem[{\citenamefont{Pomeranchuk}(1958)}]{pomeranchuk}
\bibinfo{author}{\bibfnamefont{I.}~\bibnamefont{Pomeranchuk}},
  \bibinfo{journal}{Sov. Phys. JETP} \textbf{\bibinfo{volume}{8}},
  \bibinfo{pages}{361} (\bibinfo{year}{1958}).

\bibitem[{\citenamefont{Oganesyan et~al.}(2001)\citenamefont{Oganesyan,
  Kivelson, and Fradkin}}]{eduardonematic}
\bibinfo{author}{\bibfnamefont{V.}~\bibnamefont{Oganesyan}},
  \bibinfo{author}{\bibfnamefont{S.~A.} \bibnamefont{Kivelson}},
  \bibnamefont{and} \bibinfo{author}{\bibfnamefont{E.}~\bibnamefont{Fradkin}},
  \bibinfo{journal}{Phys. Rev. B} \textbf{\bibinfo{volume}{64}},
  \bibinfo{pages}{195109} (\bibinfo{year}{2001}).

\bibitem[{\citenamefont{Barci and Oxman}(2003)}]{barci}
\bibinfo{author}{\bibfnamefont{D.~G.} \bibnamefont{Barci}} \bibnamefont{and}
  \bibinfo{author}{\bibfnamefont{L.~E.} \bibnamefont{Oxman}},
  \bibinfo{journal}{Phys. Rev. B.} \textbf{\bibinfo{volume}{67}},
  \bibinfo{pages}{205108} (\bibinfo{year}{2003}).

\bibitem[{\citenamefont{Barci and Fradkin}(2002)}]{qhe_2}
\bibinfo{author}{\bibfnamefont{D.~G.} \bibnamefont{Barci}} \bibnamefont{and}
  \bibinfo{author}{\bibfnamefont{E.}~\bibnamefont{Fradkin}},
  \bibinfo{journal}{Phys. Rev. B.} \textbf{\bibinfo{volume}{65}},
  \bibinfo{pages}{245320} (\bibinfo{year}{2002}).

\bibitem[{\citenamefont{Barci et~al.}(2002)\citenamefont{Barci, Fradkin,
  Kivelson, and Oganesyan}}]{qhe_1}
\bibinfo{author}{\bibfnamefont{D.~G.} \bibnamefont{Barci}},
  \bibinfo{author}{\bibfnamefont{E.}~\bibnamefont{Fradkin}},
  \bibinfo{author}{\bibfnamefont{S.~A.} \bibnamefont{Kivelson}},
  \bibnamefont{and}
  \bibinfo{author}{\bibfnamefont{V.}~\bibnamefont{Oganesyan}},
  \bibinfo{journal}{Phys. Rev. B.} \textbf{\bibinfo{volume}{65}},
  \bibinfo{pages}{245319} (\bibinfo{year}{2002}).

\bibitem[{\citenamefont{Honerkamp}(2005)}]{honerkamp}
\bibinfo{author}{\bibfnamefont{C.}~\bibnamefont{Honerkamp}},
  \bibinfo{journal}{Phys. Rev. B.} \textbf{\bibinfo{volume}{72}},
  \bibinfo{pages}{115103} (\bibinfo{year}{2005}).

\bibitem[{\citenamefont{Hankevych et~al.}(2002)\citenamefont{Hankevych, Grote,
  and Wegner}}]{wegner}
\bibinfo{author}{\bibfnamefont{V.}~\bibnamefont{Hankevych}},
  \bibinfo{author}{\bibfnamefont{I.}~\bibnamefont{Grote}}, \bibnamefont{and}
  \bibinfo{author}{\bibfnamefont{F.}~\bibnamefont{Wegner}},
  \bibinfo{journal}{Phys. Rev. B} \textbf{\bibinfo{volume}{66}},
  \bibinfo{pages}{094516} (\bibinfo{year}{2002}).

\bibitem[{\citenamefont{Varma and Zhu}()}]{varma}
\bibinfo{author}{\bibfnamefont{C.~M.} \bibnamefont{Varma}} \bibnamefont{and}
  \bibinfo{author}{\bibfnamefont{L.}~\bibnamefont{Zhu}},
  \bibinfo{note}{cond-mat/0502344}.

\bibitem[{\citenamefont{C.M.Varma}(2005)}]{varma_s}
\bibinfo{author}{\bibnamefont{C.M.Varma}}, \bibinfo{journal}{Phil. Mag.}
  \textbf{\bibinfo{volume}{85}}, \bibinfo{pages}{1650} (\bibinfo{year}{2005}).

\bibitem[{\citenamefont{Wu and Zhang}(2004)}]{congjun}
\bibinfo{author}{\bibfnamefont{C.}~\bibnamefont{Wu}} \bibnamefont{and}
  \bibinfo{author}{\bibfnamefont{S.-C.} \bibnamefont{Zhang}},
  \bibinfo{journal}{Phys. Rev. Lett.} \textbf{\bibinfo{volume}{93}},
  \bibinfo{pages}{036403} (\bibinfo{year}{2004}).

\bibitem[{\citenamefont{Murthy et~al.}(2004)\citenamefont{Murthy, Shankar,
  Herman, and Mathur}}]{murthy}
\bibinfo{author}{\bibfnamefont{G.}~\bibnamefont{Murthy}},
  \bibinfo{author}{\bibfnamefont{R.}~\bibnamefont{Shankar}},
  \bibinfo{author}{\bibfnamefont{D.}~\bibnamefont{Herman}}, \bibnamefont{and}
  \bibinfo{author}{\bibfnamefont{H.}~\bibnamefont{Mathur}},
  \bibinfo{journal}{Phys. Rev. B} \textbf{\bibinfo{volume}{69}},
  \bibinfo{pages}{075321} (\bibinfo{year}{2004}).

\bibitem[{\citenamefont{Metzner et~al.}(2003)\citenamefont{Metzner, Rohe, and
  Andergassen}}]{metznersoft}
\bibinfo{author}{\bibfnamefont{W.}~\bibnamefont{Metzner}},
  \bibinfo{author}{\bibfnamefont{D.}~\bibnamefont{Rohe}}, \bibnamefont{and}
  \bibinfo{author}{\bibfnamefont{S.}~\bibnamefont{Andergassen}},
  \bibinfo{journal}{Phys. Rev. Lett.} \textbf{\bibinfo{volume}{91}},
  \bibinfo{pages}{066402} (\bibinfo{year}{2003}).

\bibitem[{\citenamefont{Yang}()}]{kunyang}
\bibinfo{author}{\bibfnamefont{K.}~\bibnamefont{Yang}},
  \bibinfo{note}{cond-mat/0502270}.

\bibitem[{\citenamefont{Balibar}(2002)}]{balibarreview}
\bibinfo{author}{\bibfnamefont{S.}~\bibnamefont{Balibar}}, \bibinfo{journal}{J.
  Low. Temp. Phys.} \textbf{\bibinfo{volume}{129}}, \bibinfo{pages}{363}
  (\bibinfo{year}{2002}).

\bibitem[{voi()}]{voitlong}
\emph{\bibinfo{title}{{\rm For a review see, for instance, Johannes Voit, Rep.
  Progr. Phys. {\bf 58}, 97 (1995) and references therein.}}}

\bibitem[{\citenamefont{Haldane}(1994)}]{haldane}
\bibinfo{author}{\bibfnamefont{F.~D.~M.} \bibnamefont{Haldane}},
  \emph{\bibinfo{title}{Proceedings of the International School of Physics
  "Enrico Fermi", Course CXXI "Perspectives in Many-Particle Physics"}}
  (\bibinfo{publisher}{North-Holland}, \bibinfo{year}{1994}).

\bibitem[{\citenamefont{Houghton and Marston}(1993)}]{marston_1}
\bibinfo{author}{\bibfnamefont{A.}~\bibnamefont{Houghton}} \bibnamefont{and}
  \bibinfo{author}{\bibfnamefont{J.~B.} \bibnamefont{Marston}},
  \bibinfo{journal}{Phys.Rev.B} \textbf{\bibinfo{volume}{48}},
  \bibinfo{pages}{7790} (\bibinfo{year}{1993}).

\bibitem[{\citenamefont{Kwon et~al.}(1995)\citenamefont{Kwon, Houghton, and
  Marston}}]{marstonreview}
\bibinfo{author}{\bibfnamefont{H.-J.} \bibnamefont{Kwon}},
  \bibinfo{author}{\bibfnamefont{A.}~\bibnamefont{Houghton}}, \bibnamefont{and}
  \bibinfo{author}{\bibfnamefont{J.~B.} \bibnamefont{Marston}},
  \bibinfo{journal}{Phys.Rev.B} \textbf{\bibinfo{volume}{52}},
  \bibinfo{pages}{8002} (\bibinfo{year}{1995}).

\bibitem[{\citenamefont{{Castro Neto} and Fradkin}(1994)}]{antonioboson1}
\bibinfo{author}{\bibfnamefont{A.~H.} \bibnamefont{{Castro Neto}}}
  \bibnamefont{and} \bibinfo{author}{\bibfnamefont{E.}~\bibnamefont{Fradkin}},
  \bibinfo{journal}{Phys. Rev. B.} \textbf{\bibinfo{volume}{49}},
  \bibinfo{pages}{10877} (\bibinfo{year}{1994}).

\bibitem[{\citenamefont{{Castro Neto} and Fradkin}(1995)}]{antonioboson2}
\bibinfo{author}{\bibfnamefont{A.~H.} \bibnamefont{{Castro Neto}}}
  \bibnamefont{and} \bibinfo{author}{\bibfnamefont{E.}~\bibnamefont{Fradkin}},
  \bibinfo{journal}{Phys. Rev. B.} \textbf{\bibinfo{volume}{51}},
  \bibinfo{pages}{4084} (\bibinfo{year}{1995}).

\bibitem[{\citenamefont{Kopietz and Sch\"onhammer}(1996)}]{kopietz}
\bibinfo{author}{\bibfnamefont{P.}~\bibnamefont{Kopietz}} \bibnamefont{and}
  \bibinfo{author}{\bibfnamefont{K.}~\bibnamefont{Sch\"onhammer}},
  \bibinfo{journal}{Zeitschrift f\"ur Physik B Condensed Matter}
  \textbf{\bibinfo{volume}{100}}, \bibinfo{pages}{259} (\bibinfo{year}{1996}).

\bibitem[{\citenamefont{Shankar}(1994)}]{shankarrg}
\bibinfo{author}{\bibfnamefont{R.}~\bibnamefont{Shankar}},
  \bibinfo{journal}{Rev. Mod. Phys.} \textbf{\bibinfo{volume}{66}},
  \bibinfo{pages}{129} (\bibinfo{year}{1994}).

\bibitem[{\citenamefont{Hertz}(1976)}]{Hertz}
\bibinfo{author}{\bibfnamefont{J.}~\bibnamefont{Hertz}},
  \bibinfo{journal}{Phys. Rev. B} \textbf{\bibinfo{volume}{14}},
  \bibinfo{pages}{1165} (\bibinfo{year}{1976}).

\bibitem[{\citenamefont{Continentino et~al.}(1989)\citenamefont{Continentino,
  Japiassu, and Troper}}]{Continentino}
\bibinfo{author}{\bibfnamefont{M.~A.} \bibnamefont{Continentino}},
  \bibinfo{author}{\bibfnamefont{G.~M.} \bibnamefont{Japiassu}},
  \bibnamefont{and} \bibinfo{author}{\bibfnamefont{A.}~\bibnamefont{Troper}},
  \bibinfo{journal}{Phys. Rev. B} \textbf{\bibinfo{volume}{39}},
  \bibinfo{pages}{9734} (\bibinfo{year}{1989}).

\bibitem[{\citenamefont{Millis}(1993)}]{Millis}
\bibinfo{author}{\bibfnamefont{A.}~\bibnamefont{Millis}},
  \bibinfo{journal}{Phys. Rev. B.} \textbf{\bibinfo{volume}{48}},
  \bibinfo{pages}{7183} (\bibinfo{year}{1993}).

\bibitem[{\citenamefont{Nozieres and Pines}(1966)}]{pinesnoz}
\bibinfo{author}{\bibfnamefont{P.}~\bibnamefont{Nozieres}} \bibnamefont{and}
  \bibinfo{author}{\bibfnamefont{D.}~\bibnamefont{Pines}},
  \emph{\bibinfo{title}{The Theory of Quantum Liquids}}
  (\bibinfo{publisher}{Perseus Books}, \bibinfo{year}{1966}).

\bibitem[{\citenamefont{Feynman and Vernon}(1963)}]{feynman}
\bibinfo{author}{\bibfnamefont{R.}~\bibnamefont{Feynman}} \bibnamefont{and}
  \bibinfo{author}{\bibfnamefont{F.}~\bibnamefont{Vernon}},
  \bibinfo{journal}{Annals of Physics} \textbf{\bibinfo{volume}{24}},
  \bibinfo{pages}{118} (\bibinfo{year}{1963}).

\bibitem[{\citenamefont{Caldeira and Leggett}(1983)}]{caldeiralegget}
\bibinfo{author}{\bibfnamefont{A.}~\bibnamefont{Caldeira}} \bibnamefont{and}
  \bibinfo{author}{\bibfnamefont{A.}~\bibnamefont{Leggett}},
  \bibinfo{journal}{Physica A} \textbf{\bibinfo{volume}{121}},
  \bibinfo{pages}{587} (\bibinfo{year}{1983}).

\bibitem[{\citenamefont{Baym and Pethick}(1991)}]{baymp}
\bibinfo{author}{\bibfnamefont{G.}~\bibnamefont{Baym}} \bibnamefont{and}
  \bibinfo{author}{\bibfnamefont{C.}~\bibnamefont{Pethick}},
  \emph{\bibinfo{title}{Landau Fermi-Liquid Theory}} (\bibinfo{publisher}{John
  Wiley \& Sons}, \bibinfo{year}{1991}).

\bibitem[{not({\natexlab{a}})}]{note}
\emph{\bibinfo{title}{{\rm The same results can be obtained directly from the
  multidimensional bosonization of the theory
  \cite{antonioboson1,antonioboson2}}}}.

\bibitem[{\citenamefont{Negele and Orland}(1998)}]{negeleo}
\bibinfo{author}{\bibfnamefont{J.~W.} \bibnamefont{Negele}} \bibnamefont{and}
  \bibinfo{author}{\bibfnamefont{H.}~\bibnamefont{Orland}},
  \emph{\bibinfo{title}{Quantum Many-Particle Systems}}
  (\bibinfo{publisher}{Westview Press}, \bibinfo{year}{1998}).

\bibitem[{\citenamefont{Pethick}(1969)}]{pethickfinite}
\bibinfo{author}{\bibfnamefont{C.~J.} \bibnamefont{Pethick}},
  \bibinfo{journal}{Phys. Rev.} \textbf{\bibinfo{volume}{177}},
  \bibinfo{pages}{391} (\bibinfo{year}{1969}).

\bibitem[{\citenamefont{Chubukov et~al.}(2003)\citenamefont{Chubukov,
  Finkelstein, Haslinger, and Morr}}]{chubukov3}
\bibinfo{author}{\bibfnamefont{A.~V.} \bibnamefont{Chubukov}},
  \bibinfo{author}{\bibfnamefont{A.~M.} \bibnamefont{Finkelstein}},
  \bibinfo{author}{\bibfnamefont{R.}~\bibnamefont{Haslinger}},
  \bibnamefont{and} \bibinfo{author}{\bibfnamefont{D.~K.} \bibnamefont{Morr}},
  \bibinfo{journal}{Phys. Rev. Lett.} \textbf{\bibinfo{volume}{90}},
  \bibinfo{pages}{077002} (\bibinfo{year}{2003}).

\bibitem[{\citenamefont{Chubukov}(2005)}]{chubukovself}
\bibinfo{author}{\bibfnamefont{A.~V.} \bibnamefont{Chubukov}},
  \bibinfo{journal}{Phys. Rev. B} \textbf{\bibinfo{volume}{71}},
  \bibinfo{pages}{245123} (\bibinfo{year}{2005}).

\bibitem[{\citenamefont{Altshuler et~al.}(1994)\citenamefont{Altshuler, Ioffe,
  and Millis}}]{altshulerioffemillis2}
\bibinfo{author}{\bibfnamefont{B.~L.} \bibnamefont{Altshuler}},
  \bibinfo{author}{\bibfnamefont{L.~B.} \bibnamefont{Ioffe}}, \bibnamefont{and}
  \bibinfo{author}{\bibfnamefont{A.~J.} \bibnamefont{Millis}},
  \bibinfo{journal}{Phys. Rev. B.} \textbf{\bibinfo{volume}{50}},
  \bibinfo{pages}{14048} (\bibinfo{year}{1994}).

\bibitem[{\citenamefont{Chubukov et~al.}(2004)\citenamefont{Chubukov, Pepin,
  and Rech}}]{chubukov2}
\bibinfo{author}{\bibfnamefont{A.~V.} \bibnamefont{Chubukov}},
  \bibinfo{author}{\bibfnamefont{C.}~\bibnamefont{Pepin}}, \bibnamefont{and}
  \bibinfo{author}{\bibfnamefont{J.}~\bibnamefont{Rech}},
  \bibinfo{journal}{Phys. Rev. Lett.} \textbf{\bibinfo{volume}{92}},
  \bibinfo{pages}{147003} (\bibinfo{year}{2004}).

\bibitem[{not({\natexlab{b}})}]{notenew}
\emph{\bibinfo{title}{{\rm After our paper was submitted we became aware of
  other work with similar results\cite{eduardonew}}.}}

\bibitem[{\citenamefont{Lawler et~al.}(2005)\citenamefont{Lawler, Fernandez,
  Barci, Fradkin, and Oxman}}]{eduardonew}
\bibinfo{author}{\bibfnamefont{M.~J.} \bibnamefont{Lawler}},
  \bibinfo{author}{\bibfnamefont{V.}~\bibnamefont{Fernandez}},
  \bibinfo{author}{\bibfnamefont{D.~G.} \bibnamefont{Barci}},
  \bibinfo{author}{\bibfnamefont{E.}~\bibnamefont{Fradkin}}, \bibnamefont{and}
  \bibinfo{author}{\bibfnamefont{L.}~\bibnamefont{Oxman}}
  (\bibinfo{year}{2005}), \bibinfo{note}{cond-mat/0508747}.

\bibitem[{\citenamefont{Kee et~al.}(2003)\citenamefont{Kee, Kim, and
  Chung}}]{kee}
\bibinfo{author}{\bibfnamefont{H.-Y.} \bibnamefont{Kee}},
  \bibinfo{author}{\bibfnamefont{E.~H.} \bibnamefont{Kim}}, \bibnamefont{and}
  \bibinfo{author}{\bibfnamefont{C.-H.} \bibnamefont{Chung}},
  \bibinfo{journal}{Phys. Rev. B} \textbf{\bibinfo{volume}{58}},
  \bibinfo{pages}{245109} (\bibinfo{year}{2003}).

\bibitem[{\citenamefont{Khavkine et~al.}(2004)\citenamefont{Khavkine, Chung,
  Oganesyan, and Kee}}]{khavkine}
\bibinfo{author}{\bibfnamefont{I.}~\bibnamefont{Khavkine}},
  \bibinfo{author}{\bibfnamefont{C.-H.} \bibnamefont{Chung}},
  \bibinfo{author}{\bibfnamefont{V.}~\bibnamefont{Oganesyan}},
  \bibnamefont{and} \bibinfo{author}{\bibfnamefont{H.-Y.} \bibnamefont{Kee}},
  \bibinfo{journal}{Phys. Rev. B} \textbf{\bibinfo{volume}{70}},
  \bibinfo{pages}{155110} (\bibinfo{year}{2004}).

\end{thebibliography}

\end{document}